\documentstyle[aps]{revtex}
\input epsf

\begin{document}


\draft

\title{Pseudo-spectral apparent horizon finders: an efficient new algorithm}

\author{Carsten Gundlach
\thanks{Electronic address: gundlach@aei-potsdam.mpg.de}}
\address{Max-Planck-Institut f\"ur Gravitationsphysik,
Albert-Einstein-Institut, Schlaatzweg 1, 14473 Potsdam, Germany}

\date{23 July 1997, revised 6 October 1997}

\maketitle


\begin{abstract}
  

We review the problem of finding an apparent horizon in Cauchy data
$(\Sigma,g_{ab},K_{ab})$ in three space dimensions without
symmetries. We describe a family of algorithms which includes the
pseudo-spectral apparent horizon finder of Nakamura et al. and the
curvature flow method proposed by Tod as special cases. We suggest
that other algorithms in the family may combine the speed of the
former with the robustness of the latter.  A numerical implementation
for Cauchy data given on a grid in Cartesian coordinates is described,
and tested on Brill-Lindquist and Kerr initial data. The new algorithm
appears faster and more robust than previous ones.

\end{abstract}



\section{Introduction}


An important task in numerical relativity is locating black holes in
numerically generated spacetimes, both for technical purposes and for
extracting physical information. A black hole is a region of spacetime
out of which no null geodesics escape to infinity. The boundary of the
black hole, the event horizon, is formed by those outward-going,
future-directed null geodesics which neither fall into the singularity
nor escape to null infinity. The event horizon contains important
geometric information about the spacetime. It is a global construction
and can in principle only be determined when the entire spacetime is
known.  In practice, one can obtain a good approximation to the event
horizon within a finite spacetime region, once the black hole has
settled down to a stationary state. By definition, the event horizon
repels future-directed null geodesics, but attracts past ones. One can
then evolve past-directed null geodesics back through the spacetime,
and find the event horizon as the surface to which they are attracted
\cite{EHF}. 

Locating black holes is crucial in numerical relativity also for a
technical reason: Spacetime slicings which avoid black holes rapidly
become singular. Instead one would like to excise a spacetime region
just inside the event horizon from the numerical domain during the
numerical evolution, using the fact that it cannot influence events
outside the black hole. During the time evolution, however, one does
not yet know where the event horizon is. Instead one needs to use the
poor man's event horizon, the apparent horizon.

An apparent horizon (AH) is defined within a single time slice, or
spacelike hypersurface $\Sigma$, namely as a smooth embedded 2-surface
whose outgoing normal null geodesics have zero expansion. There may be
one such surface enclosing another one, in which case the outermost
one is the apparent horizon.  If one combines the apparent horizon on
each time slice into a 3-dimensional surface, this world tube will
depend on the slicing, and can be discontinuous.  Nevertheless one can
show that if an apparent horizon exists on a given time slice, it must
be inside a black hole \cite{Wald}. The converse is not true: there
are slicings of black hole spacetimes without any apparent horizons
\cite{noAH}. For numerical purposes one simply hopes that this case is
unusual and that the apparent horizon gives a reasonable indication of
the event horizon.

A wide variety of numerical algorithms for finding AHs have been
explored or suggested.  For the purpose of excising the black hole
region, one needs to find the apparent horizon frequently, perhaps at
each time step. When two black holes collide, a new AH enveloping the
two separate ones appears suddenly. Therefore the main requirements
are speed and finding the AH from scratch, without a good initial
guess. Precision is less important for black hole excision, although a
safe error estimate is, so that one can be sure not to excise too much
and inject unphysical boundary conditions.

In spherical symmetry, the AH problem reduces to an algebraic
equation. In axisymmetry, it reduces to an ordinary differential
equation with periodic boundary conditions. In this paper we shall be
concerned exclusively with the 3-dimensional problem without any
symmetries, either continuous or discrete, where one deals with a
highly nonlinear elliptic problem on a closed 2-surface. In practice
this will always be the 2-sphere, or several disconnected 2-spheres
\cite{Gibbons}, which can be treated separately.  

All 3D AH finder algorithms proposed so far can be classified
according to a few key choices, which can be made independently one
from another. How are candidate AHs represented? One can parameterize
an embedded 2-surface either by introducing coordinates on it, or as a
level set of a function on the 3-dimensional slice in which it is
embedded.  How is the curvature of the candidate AH calculated?  One
can discretize the necessary spatial derivatives by finite
differencing, finite elements, or pseudo-spectral methods.  A third
fundamental choice is between solving the elliptic problem directly,
or converting it into a parabolic problem, in which the solution of
the elliptic problem is approached during an evolution in an
unphysical time parameter.  The distinction between these last two
approaches is not sharp in practice. On the one hand one always solves
a nonlinear elliptic problem by iteration. On the other, numerical
implementation of any parabolic approach requires an implicit ``time''
step for stability, thus posing a new elliptic problem that becomes
equivalent to the original one in the limit of an infinitely large
time step. We now discuss previous AH finders in terms of these choices.

Nakamura, Koshima and Oohara \cite{NKO} represent the AH in spherical
coordinates as $r=h(\theta,\varphi)$.  We note that this requires the
surface to be a 2-sphere, and star-shaped (convex) around the point
$r=0$. The shape function $h(\theta,\varphi)$ is expanded in spherical
harmonics. This spectral decomposition is used to calculate the
derivatives of $h$ in formulating the elliptic problem. The
orthonormality and completeness of the spherical harmonics is used to
subtract the linear elliptic operator $L^2$ from the nonlinear
elliptic problem and invert it. This gives rise to an iteration
prescription. We shall see that this iteration can also be described
as the discretization in unphysical time of a parabolic problem. It
remains unclear from
\cite{NKO} by what method the spectral decomposition back and forth is
carried out for Cauchy data which are only known in numerical form and
on a grid.  The Nakamura {\it et al.}  algorithm has been
independently coded and tested, and extended in various directions by
Kemball and Bishop \cite{KB}. They report exponential convergence,
good robustness, and high precision unless the point $r=0$ is close to
the AH.

Tod \cite{Tod} has proposed a geometrically defined flow under which a
trial 2-surface evolves to the AH. For time-symmetric slices, the AH
problem reduces to that of finding a minimal surface, and Tod's
prescription to mean curvature flow. This is well-known to converge to
minimal surfaces. On non time-symmetric slices, only lower order terms
are added to the problem, so that one may hope that Tod's flow also
converges for such data in practice. Tod's algorithm is parabolic,
without specifying how the surface is represented or differenced.
Tod's algorithm has been implemented numerically by Bernstein
\cite{Bernstein} using finite differencing in coordinates introduced
on the surface. He discusses stable extrinsic algorithms for parabolic
problems, and reports good results in axisymmetry using one of them,
but technical problems to do with finite differencing on the sphere in
the general case.  Pasch \cite{Pasch} has implemented mean curvature
flow representing the test surface as a level set. This allows for a
change of topology during the evolution. He has successfully tested
the algorithm using Brill-Lindquist data for 1, 2 or 3 black holes,
using a fast implicit time evolution package, and finite differencing
on a Cartesian grid in the embedding space.

Thornburg \cite{Thornburg} attacks the elliptic problem directly using
finite differencing on a square $(\theta,\varphi)$ grid, and Newton's
method to solve the discretized equations. He calculates the Jacobian
required for Newton's method by first linearizing the differential
equations, then finite differencing the result. This is more efficient
than numerical differentiation. He finds high precision results, but a
nonlinear instability in high-frequency modes.  Huq \cite{Huq} has
extended Newton's method to data without symmetries in Cartesian-type
coordinates.

The NCSA/WashU algorithm \cite{NCSA1,NCSA2} uses the parameterization
$r=h(\theta,\varphi)$, and a spectral decomposition to parameterize
$h$ and calculate its derivatives. The discretized elliptic problem is
solved by applying a standard minimization algorithm to the sum of
$H^2$ over all surface points. The spectral basis is not required to
be orthonormal for this purpose.  Baumgarte {\it et al.}
\cite{Cornell} have implemented the NCSA/WashU algorithm
independently, with the difference that they use the true spherical
harmonics as a basis. Both algorithms locate points on the 2-surface
on a square $(\theta,\varphi)$ grid, interpolating data from the
3-dimensional Cartesian grid used in the 3+1 time evolution.

In this paper we review different previous algorithms in one common,
fully covariant notation. This analysis suggests to us a new algorithm
which combines essential ideas of the algorithms of Tod and Nakamura
{\it et al.} From our analysis we expect this algorithm to be as fast
as that of Nakamura {\it et al.} (and therefore much faster than
existing implementations of Tod's algorithm), while being as robust in
practice as that of Tod. We describe the details of a numerical
implementation of this algorithm, and some initial tests. The results
are encouraging. 

The paper is organized as follows.  In section II we set up the
mathematical formalism of the problem.  We begin by deriving the
differential equation that determines an apparent horizon in II.A. In
II.B we discuss different ways of parameterizing apparent horizon
candidates, that is, smooth embedded 2-surfaces, and in II.C we
provide tools for spectral methods on the 2-sphere. In section III we
review various algorithms for finding AHs, namely the pseudo-spectral
algorithm of Nakamura {\it et al.} \cite{NKO}, Jacobi's method, and
the generalized mean curvature flow suggested by Tod
\cite{Tod}. 
We then build on this review by presenting a family of algorithm which
contains the previous algorithms as limiting cases, and suggesting
that in the middle of the family there are algorithms that perform
better than the limiting members.  In section IV we describe a
numerical implementation of our proposed algorithm. In section V we
test its performance in finding apparent horizons in Brill-Lindquist
and Kerr data given in Cartesian coordinates on a grid.


\section{Mathematical preliminaries}


\subsection{The apparent horizon equation}


Here we give a brief derivation (see also \cite{York}) of the
differential equation that we try to solve in the remainder of the
paper, both to give a complete presentation of the problem and to fix
notation.  Throughout the paper, lower-case Latin indices from the
beginning of the alphabet indicate abstract index notation. Indices
from the middle of the alphabet indicate 3-dimensional tensor {\it
components}. Our signature convention is $(-+++)$.

We begin with a series of definitions.  Let $(M,{}^{(4)}g_{ab})$ be a
spacetime, and $\nabla^{(4)}_a$ the covariant derivative associated
with ${}^{(4)}g_{ab}$. (We use this notation to reserve the symbols
$g_{ab}$ and $\nabla_a$ for 3 dimensions.)  Let $\Sigma$ be a smooth
spacelike hypersurface, and let $n^a$ be the future-pointing unit
timelike normal to $\Sigma$. Then ${}^{(4)}g_{ab}$ gives rise to
Cauchy data
\begin{equation}
g_{ab} = {}^{(4)}g_{ab} + n_a n_b, \quad K_{ab} = - g_a^{\ c} \
\nabla^{(4)}_c n_b = - \nabla_a n_b,
\end{equation}
on $\Sigma$, where $g_{ab}$ is the positive definite 3-metric induced on
$\Sigma$ and $K_{ab}$ is the extrinsic curvature of
$\Sigma$. $\nabla_a$ is the covariant derivative associated with
$g_{ab}$.  Let $S$ be a closed smooth hypersurface of $\Sigma$, which
means it is two-dimensional and spacelike, and $s^a$ its unit outward
pointing normal in $\Sigma$, which is also spacelike, and normal to
$n^a$. $g_{ab}$ induces a positive definite 2-metric
\begin{equation} 
m_{ab} = g_{ab} - s_a s_b = {}^{(4)}g_{ab} + n_a n_b - s_a s_b
\end{equation}
on $S$.  Let $k^a$ be the future-pointing null geodesic congruence
whose projection on $\Sigma$ is orthogonal to $S$, that is
\begin{equation}
k^a\nabla^{(4)}_a k^b=0, \quad  k_a k^a=0, \quad m_{ab}{k^a}|_S=0.
\end{equation}
Then $k^a$ describes light rays leaving $S$ normally from the point of
view of an observer whose instantaneous simultaneity is $\Sigma$. Clearly
$k^a$ depends not only on the spacetime and on $S$, but also on
$\Sigma$.  Let $H$ be the expansion of that congruence,
\begin{equation}
H = \nabla^{(4)}_a k^a.
\end{equation}

We would like to express $H$ in terms of the Cauchy data
$(\Sigma,g_{ab},K_{ab})$, and the normal $s^a$ to $S$. The crucial
step is to note that, up to an overall factor,
\begin{equation}
{k^a}|_S=s^a + n^a.
\end{equation}
Clearly this obeys the conditions $m_{ab}k^a=0$ and $k_a k^a=0$ on
$S$. We continue $k^a$ away from $S$ by the remaining condition
$k^a\nabla^{(4)}_a k^b=0$. We also continue $s^a$ away from $S$ in
$\Sigma$ assuming that it retains unit length, but otherwise in an
arbitrary manner.  Then we have, on $S$,
\begin{eqnarray}
H && = {}^{(4)}g^{ab} \nabla^{(4)}_a k_b 
= (g^{ab}-n^an^b) \nabla^{(4)}_a k_b \cr
&& = g^{ab} 
\nabla^{(4)}_a (s_b + n_b) - (k^a-s^a)(k^b-s^b)\nabla^{(4)}_a k_b \cr 
&& = g^{ab} 
\nabla^{(4)}_a s_b + g^{ab} \nabla^{(4)}_a n_b - (k^b-s^b)[k^a \
\nabla^{(4)}_a k_b]
+s^a[k^b \nabla^{(4)}_a k_b] - s^a [s^b \nabla^{(4)}_a s_b] 
- s^a s^b \nabla^{(4)}_a n_b \cr 
\label{H}
&& = \nabla_a
s^a - K + s^a s^b K_{ab},
\end{eqnarray}
where $K={}^{(4)}g^{ab}K_{ab}=g^{ab}K_{ab}$ is the trace of the extrinsic
curvature. All terms in square brackets vanish individually by definition.

A smooth embedded closed surface with outward pointing unit normal
$s^a$ that obeys $H=0$ everywhere on $S$ is called a marginally outer
trapped surface. The outermost of such surfaces, if one or more exist
in $\Sigma$, is called the apparent horizon in $\Sigma$
\cite{Wald}. On the one hand this definition is global in $\Sigma$,
which makes finding an apparent horizon a non-trivial problem.  On the
other, it is local in time, as $H$ depends only on the Cauchy data
$(g_{ab},K_{ab})$ on a single slice $\Sigma$. If one fixes the slicing
of a given spacetime, calculates the apparent horizon on each slice,
and then combines the apparent horizons on each slice to obtain a
timelike, $2+1$ dimensional world-tube, this world-tube depends on the
slicing. This is in contrast to the event horizon, which depends
globally on the entire spacetime, but is independent of the slicing.


\subsection{Characterizing closed two-surfaces}


Before we can discuss solving the apparent horizon equation $H=0$ in
practice, we need to parameterize candidate apparent horizons, that
is, two-dimensional, smooth, closed surfaces $S$ embedded in $\Sigma$.

Let $x^i$ be coordinates on $\Sigma$. One way of parameterizing $S$ is
then to introduce coordinates $\xi^A$ on $S$ (at least locally), and
give a map $x^i=X^i(\xi^A)$. In this case, the topology of $S$ is
fixed in advance. Furthermore, different functions $X^i$ describe the
same abstract surface $S$, corresponding to a change of coordinates
$\xi^A$ on $S$.

A different way of parameterizing $S$ is as a level set
\begin{equation}
F(x^i)=0. 
\end{equation}
As long as the form of $F(x^i)$ is not restricted, this has the
advantage of $S$ allowing to have arbitrary topology. In particular,
$S$ can be disconnected. Again, many different functions $F(x^i)$
describe the same abstract surface $S$, as long as they have the one
level set $F(x^i)=0$ in common. 

It is straightforward to express $H$ as a function of $F$ and its
derivatives. The unit normal (with respect to the 3-metric $g_{ab}$) of
any level set of $F$ is
\begin{equation}
\label{s^a}
s^a = |\nabla  F|^{-1} g^{ab} \nabla_b F, \quad \text{where} \quad
|\nabla  F| \equiv \left(g^{ab} \nabla_a F \nabla_b F\right)^{1/2}.
\end{equation}
Direct substitution now gives
\begin{equation}
H 
= \left(g^{ab} - |\nabla F|^{-2} \nabla^a F \nabla^b F\right)
\left( |\nabla F|^{-1} \nabla_a \nabla_b F - K_{ab}\right) =
m^{ab} \left( |\nabla F|^{-1} \nabla_a \nabla_b F - K_{ab}\right).
\end{equation}
$H$ is therefore a quasi-linear second order differential operator
acting on $F$.  

Now we come back to the problem that different functions $F(x,y,z)$
describe the same abstract surface $S$. A possible gauge condition
would be to make $F$ harmonic with respect to a background metric
$\bar g_{ab}$, or with respect to the physical 3-metric $g_{ab}$. Then
its value everywhere depends only on its value on a suitable
two-dimensional surface, such as the boundary of the numerical domain.
Here, instead, we follow several previous authors in restricting $F$
to the form
\begin{equation}
F(x^i)=r(x^i)-h[\theta(x^i),\varphi(x^i)],
\end{equation}
where $(r,\theta,\varphi)$ are related to a set of Cartesian
coordinates $x^i$ in the usual way, namely $x=r\sin\theta\cos\varphi$,
$y=r\sin\theta\sin\varphi$ and $z=r\cos\theta$. The overall sign of
$F$ has been chosen so that $s^a$ given in (\ref{s^a}) points
outward. This parameterization is equivalent to
$X^i(\theta,\varphi)=x^i[r=h(\theta,\varphi),\theta,\varphi]$. The
obvious disadvantages of restricting $F$ to this form are that the
topology of $S$ must be $S^2$, and that $S$ must be star-shaped around
the coordinate origin $r=0$. The advantages are that surfaces $S$
correspond uniquely to functions $h$, and that we can use the natural
basis $\{Y_{lm}\}$ for expanding the function $h$.

Considered as a quasi-linear differential operator acting on $F(x^i)$,
$H$ is not elliptic in three dimensions, because one of the three
eigenvalues of $m^a_{\ b}$, the one with eigenvector $s^a$, is zero.
Considered as a differential operator in two dimensions acting on
$h(\theta,\varphi)$, it is elliptic. In this two-dimensional
interpretation it is nonlinear not only through the explicit
appearance of $\nabla_a F$ in the coefficients of $\nabla_a\nabla_bF$,
but also through its dependence on the point where the tensor fields
$g_{ab}$ and $K_{ab}$ are evaluated, which depends on $F$ itself. This
means that $g_{ab}$ and $K_{ab}$ play the same role in the apparent
horizon equation as the internal metric of a nonlinear $\sigma$-model
does in its equation of motion. Because the coefficients of the
elliptic equation contain $g_{ab}$ and $K_{ab}$ as free functions, it
appears unlikely that one can prove existence of solutions for
sufficiently general $g_{ab}$ and $K_{ab}$.


\subsection{Geometric characterization of the $L^2$ operator and
spherical harmonics}


In this subsection we introduce in geometric terms some tools that we
need later on to discuss spectral methods on the 2-sphere.  The
key idea of any pseudo-spectral method for solving a nonlinear
elliptic problem is to subtract from the nonlinear one a simple linear
elliptic operator that can be inverted explicitly by spectral
methods. In our problem, the principal part of the operator $H$ acting
on $h$ is the Laplacian with respect to the 2-dimensional metric
$m_{ab}$ induced on the surface $F=0$ by the metric $g_{ab}$. As $F=0$
is topologically a 2-sphere, a natural candidate for subtraction is
the Laplacian $L^2$ on the round 2-sphere. It can be inverted using
the spherical harmonics.

We could define $L^2$ on $\Sigma$ by first introducing spherical
coordinates $(r,\theta,\varphi)$, and then defining its action on a
scalar $f$ as the usual combination of partial derivatives with
respect to those coordinates, that is,
\begin{equation}
\label{L2standard}
L^2 f = f_{,\theta\theta} + \cot\theta f_{,\theta} + \sin^{-2}\theta
f_{,\varphi\varphi}.
\end{equation}
Setting up these coordinates also has the effect of lifting the
spherical harmonics from the 2-sphere to all of $\Sigma$, by smoothly
identifying points on different spheres $r=const.$ 

The minimal geometric structure which allows us to make the same
definitions without reference to preferred coordinates is a flat
background metric $\bar g_{ab}$ on $\Sigma$ (independent of the
physical metric $g_{ab}$), together with a preferred point $O$.  Let
the covariant derivative associated with $\bar g_{ab}$ be
$\bar\nabla_a$, and let $\bar g ^{ab}$ be the inverse of $\bar
g_{ab}$. We foliate $\Sigma$ into level surfaces of the scalar field
$r$, where $r(p)$ is the geodesic distance with respect to $\bar
g_{ab}$ between the points $p$ and $O$. The vector $r^a\equiv\bar
g^{ab}\bar\nabla_b r$ is the unit normal with respect to $\bar g_{ab}$
on the surfaces of constant $r$.  The flat metric $g_{ab}$ then
induces the metric $\bar g_{ab}-\bar\nabla_a r \bar\nabla_b r$ on the
surfaces of constant $r$. This induced metric has a constant curvature
of $r^{-2}$, so that $r^{-2}(\bar g_{ab}-\bar\nabla_a r \bar\nabla_b
r)$ is a metric of unit curvature on the 2-spheres $r=const.$ We now
define $L^2$ as the Laplacian of this 2-dimensional metric:
\begin{equation}
\label{L^2}
L^2 = r^2 (\bar g^{ab} - r^a r^b) \bar\nabla_a \bar\nabla_b - 2 r r^a
\bar\nabla_a.
\end{equation}
By direct substitution one verifies that, if $\bar g_{ab}$ is given as
\begin{equation}
\label{standcoord}
ds^2 = dr^2 + r^2(d\theta^2+\sin^2\theta d\varphi^2),
\end{equation}
this reduces to Eq. (\ref{L2standard}).  Our definition (\ref{L^2}),
however, is covariant, and can be used to define the action of $L^2$
on arbitrary tensors, and in arbitrary coordinate systems.

For our purposes we characterize the spherical harmonics $Y_{lm}$ as a
set of scalar functions on $\Sigma$ with two properties: They are
orthonormal in the sense that
\begin{equation}
\label{cond0}
\int_S Y_{lm}^{*} Y_{l'm'} d\Omega = \delta_{ll'} \delta_{mm'},
\end{equation}
where $S$ is any smooth surface that is
star-shaped around $r=0$, and 
where $d\Omega$ is the measure induced on $S$ by
$r^{-2}(\bar g_{ab} - \bar\nabla_a r \bar\nabla_b r)$. (In spherical
coordinates this reduces to the standard measure $d\Omega=\sin\theta
d\theta d\varphi$.)  From this it follows that
\begin{equation}
\label{cond1}
r^a\bar\nabla_a Y_{lm}=0.
\end{equation}
[In spherical coordinates $Y_{lm}=Y_{lm}(\theta,\varphi)$.]  We also
require that the $Y_{lm}$ are eigenfunctions of $L^2$:
\begin{equation}
\label{cond2}
L^2 Y_{lm} = -l(l+1) Y_{lm},
\end{equation}
for $l=0,1,2,\dots$ and $m=-l,\dots,l$.  We do not define $m$ as the
eigenvalue of $L_z$ ($\partial/\partial\varphi$ in spherical
coordinates), but only use it as a label on the orthonormal
basis. This leaves us free to combine $Y_{lm}$ and $Y_{l,-m}$ of the
standard complex definition to obtain a real orthonormal basis more
convenient for numerical purposes.


\section{Algorithms for solving the apparent horizon equation}


\subsection{The Nakamura {\it et al.} algorithm}


We now use our covariant notation for $L^2$ and the $Y_{lm}$ in
reviewing the algorithm of Nakamura, Kojima and Oohara \cite{NKO}
(from now on NKO) for finding an apparent horizon. NKO characterize
$S$ by $r=h(\theta,\varphi)$ in spherical coordinates, and expand $h$
in spherical harmonics:
\begin{equation}
h(\theta,\varphi) = \sum_{l=0}^{l_{\text{max}}} \sum_{m=-l}^l a_{lm}
Y_{lm}(\theta,\varphi).
\end{equation}
(A finite value of $l_{\text{max}}$ is required in any numerical
implementation.)  We begin our description of the algorithm with the
trivial observation that $H=0$ is equivalent to
\begin{equation}
\rho H + L^2 h =  L^2 h,
\end{equation}
where $\rho$ is any strictly positive function. In the NKO algorithm,
the weight function $\rho$ is specified by demanding that the
coefficient of the partial derivative $h_{,\theta\theta}$ cancels in
the combination $\rho H + L^2 h$.  (The notation $\rho$ is ours, not
that of NKO. We introduce it here because we want to consider other
choices of $\rho$ later on.)  Integrating over the $Y_{lm}$ and using
(\ref{cond0}) and (\ref{cond2}), we obtain
\begin{equation}
\int_S Y_{lm}^{*} (\rho H + L^2 h) d\Omega = - l(l+1)a_{lm}.
\end{equation}
NKO  now use this equation in an iteration procedure, 
$\{a_{lm}\}^{(n)} \to \{a_{lm}\}^{(n+1)}$, where $(n)$
labels iteration steps, of the form
\begin{equation}
\label{l>0}
a_{lm}^{(n+1)} = - {1\over l(l+1)} 
\int_S Y_{lm}^{*} (\rho H + L^2 h)^{(n)} d\Omega,
\end{equation}
where the right-hand side is evaluated from the $\{a_{lm}\}^{(n)}$. As
this formula does not cover $a_{00}$, $a_{00}$ is determined at each
iteration step by solving 
\begin{equation}
\label{a00}
\int_S (\rho H + L^2 h) d\Omega = 0
\end{equation}
for $a_{00}$ (note $Y_{00}^{*}=const.$) 

We now try to understand what makes the NKO method work.  For this
purpose we express $H(h)$ in terms of the flat
background derivation $\bar\nabla_a$:
\begin{equation}
\label{H(F)}
H = 
(g^{ab}-s^a s^b)
\left\{ 
|\nabla F|^{-1} 
\left[\bar\nabla_a \bar\nabla_b F - {1\over2} g^{cd} 
(\bar\nabla_a g_{cb} + \bar\nabla_b g_{ac} - \bar\nabla_c g_{ab})
\bar\nabla_d F \right] 
- K_{ab} \right\},
\end{equation}
where $F\equiv r-h(\theta,\varphi)$, and $s^a$ is defined by Eq.
(\ref{s^a}). Because $r^a\bar\nabla_a h=0$ by definition, we have
\begin{equation}
\label{L2h}
L^2 h = r^2 (\bar g^{ab}-r^a r^b) \bar\nabla_a \bar\nabla_b h.
\end{equation}
Putting Eqs. (\ref{H(F)}) and (\ref{L2h}) together, and keeping in
mind that $F=r-h$, we obtain 
\begin{equation}
\rho H + L^2 h = M^{ab} \bar\nabla_a \bar\nabla_b h + W,
\end{equation}
where $M^{ab}$ and $W$ depend explicitly on first derivatives of $h$,
and implicitly on $h$ through the point in $\Sigma$ where $g_{ab}$ and
$K_{ab}$ are evaluated. We have quietly assumed that $\rho$ does not
depend on second or higher derivatives of $h$, so that $\rho H+L^2h$,
like $H$ itself, acts on $h$ as a quasi-linear second-order
differential operator. This is indeed the case for the
$\rho$ of NKO and the other choices we explore later on. The principal
symbol $M^{ab}$ is
\begin{equation}
M^{ab}= - \rho |\nabla  F|^{-1}
(g^{ab}-s^a s^b) 
+ r^2 (\bar g^{ab}-r^a r^b).
\end{equation}
The principal symbol of a quasi-linear differential operator does not
depend on the choice of derivation, here $\bar\nabla_a$. We can verify
this for the case at hand.

We see that $M^{ab}$ is a difference between two projectors: the first
one onto surfaces of constant $F$, and with respect to the physical
metric $g_{ab}$, the other onto surfaces of constant coordinate $r$,
and with the respect to the background metric $\bar g_{ab}$. In the
trivial case where $g_{ab}$ is conformally related to $\bar g_{ab}$
(conformally flat), and where surfaces of constant $F$ coincide with
surfaces of constant $r$ (spherical symmetry), one can choose $\rho$
so as to make the entire tensor $M^{ab}$ vanish. In general, one can
impose only one condition on its six components. The choice of NKO
is, in our notation, 
\begin{equation}
M^{\theta\theta}=0.
\end{equation}
We prefer a coordinate-independent choice, and impose 
\begin{equation}
M^{ab}(\bar g_{ab}-\bar\nabla_a r \bar\nabla_b r) = 0.
\end{equation}
The motivation of either choice is to cancel, as far as possible, that
part of $M^{ab}\bar\nabla_a \bar\nabla_b$ which looks like $L^2$. Our
choice does not introduce a preferred direction within
the tangent space of $S$, which may be an esthetic more than a
practical advantage.  Solving our condition for $\rho$, we obtain
\begin{equation}
\label{rho}
\rho = 2 r^2|\nabla  F| \left[(g^{ab}-s^a s^b) (\bar g_{ab}-\bar\nabla_a r
\bar\nabla_b r)\right]^{-1} \equiv |\nabla  F| \sigma,
\end{equation}
where the second equation defines $\sigma$.  

Now we recognize an important ingredient of the NKO algorithm, its
smoothing property. Putting the individual components $a_{lm}$ back
together again, we can write (\ref{l>0}) as
\begin{equation}
\label{30}
h^{(n+1)} = (L^2)^{-1} (\rho H + L^2) h^{(n)}.
\end{equation}
(This is only formal because of the special role of $a_{00}$: $L^2$
does not have an inverse.)  Any iterative algorithm for solving an elliptic
problem runs the danger of being unstable to the growth of
high-frequency numerical noise.  Whereas $H$ acts on $h$ as a
second-order differential operator, thus increasing unsmoothness,
$\rho H + L^2$ has the $L^2$ part taken out, and therefore creates
less high-frequency noise. Moreover, $(L^2)^{-1}$ acts as a smoothing
operator. One therefore expects $h^{(n+1)}$ to be smoother than
$h^{(n)}$. This is a necessary property for any iterative algorithm
that can converge from a rough initial guess without blowing up
through high-frequency noise on the way.


\subsection{Jacobi's method, and stability}


In order to see how the NKO algorithm is related to other algorithms,
we rewrite (\ref{30}) once more, as
\begin{equation}
h^{(n+1)} - h^{(n)} = (L^2)^{-1}(\rho H)^{(n)}.
\end{equation}
It is now tempting to go from the discrete algorithm to a continuous
flow in an unphysical ``time'' parameter $\lambda$:
\begin{equation}
{\partial h(\theta,\varphi,\lambda) \over \partial \lambda}
= (L^2)^{-1} \rho H(h).
\end{equation}
The NKO algorithm proper, namely
\begin{equation}
\label{NKOcomponents}
a_{lm}^{(n+1)} - a_{lm}^{(n)}  = - {1\over l(l+1)}
(\rho H)_{lm}^{(n)}, \quad l>0,
\end{equation}
is formally recovered from this differential equation by
forward-differencing it with respect to $\lambda$, with a step size
$\Delta\lambda=1$, and inverting $L^2$ by the pseudo-spectral method
(we again disregard the special role of $a_{00}$). Other differencing
methods, such as centered differencing, and using a different ``time''
step, give rise to obvious alternative algorithms. Some of these have
been examined by Kemball and Bishop \cite{KB}. Kemball and Bishop also
consider different methods of enforcing the constraint (\ref{a00}) on
$a_{00}$ and of coupling it to the iteration method for the other
$a_{lm}$.

Any flow method can be considered as an example of Jacobi's
method. This is the recipe of solving an elliptic equation $E(h)=0$ by
transforming it into a parabolic equation $\partial h/\partial \lambda
= E(h)$. If $E$ is the Laplace operator, then the resulting equation
is the heat equation, and Jacobi's method is known to converge. As $H$
acting on $h$ resembles $-L^2$, one might try the flow
\begin{equation}
\label{Jacobi}
{\partial h\over \partial \lambda} = - H(h).
\end{equation}
We have implemented this numerically in the pseudo-spectral framework
and find empirically that its high frequency noise blows up unless one
chooses a very small step size.  

The origin of this instability is clear from the analogy with the heat
equation. The heat equation on $S^2$ is
\begin{equation}
  f_{,\lambda} = L^2 f.
\end{equation}
We decompose $f$ into spherical harmonics, as
$f(\theta,\varphi,\lambda) = \sum f_{lm}(\lambda)
Y_{lm}(\theta,\varphi)$.  For the spectral components we obtain
$df_{lm}/d\lambda = - l(l+1) f_{lm}$. All spectral components decrease
exponentially.  Discretizing this equation in time, however, for
example by forward differencing, we obtain$f_{lm}^{(n+1)} = f_{lm}^{(n)} -
\Delta \lambda\, l(l+1) f_{lm}^{(n)}$. This is stable only for
$|1-\Delta \lambda l(l+1)| < 1$, that is for $\Delta \lambda < 2 /
l(l+1)$. In all explicit methods, the time step is limited by order of
magnitude to
\begin{equation}
\Delta \lambda \lesssim
{1\over l_{\text{max}}(l_{\text{max}}+1)}.
\end{equation}
The same limit arises if one discretizes $L^2$ by finite differencing,
where it takes the form $
\Delta \lambda \lesssim
(\Delta \theta)^2 \simeq (\Delta \varphi)^2.  $ Similar stability
limits exist for all parabolic equations.  The NKO algorithm does not
have this instability problem. It replaces $H=0$ by $(L^2)^{-1}(\rho
H)=0$ as the elliptic problem to be solved, and clearly $(L^2)^{-1}$
acts a smoothing operator that keeps high frequency noise down. An
appropriate choice of $\rho$ makes this even more effective by making
$\rho H$ as similar to $-L^2$ as possible.


\subsection{Mean curvature flow}


From considering an iterative approach as the discretization of a flow
on the space of surfaces, one is led to the generalized mean curvature flow
algorithm of Tod \cite{Tod} and other geometrically motivated
flows. Tod proposes deforming a trial surface $S$ embedded in $\Sigma$
by means of the flow
\begin{equation}
\label{Todeqn}
\left({\partial \over \partial \lambda}\right)^a = - s^a H,
\end{equation}
where $s^a$ is again the outward-pointing normal to $S$. [Tod uses the
notation $dx^i/d\lambda$ for the left-hand side, but we wanted to
stress here that $(\partial/ \partial \lambda)^a$ is a vector field
and independent of coordinates.] For time-symmetric Cauchy data,
$K_{ab}=0$, we have $H=\nabla_a s^a$, which is simply the trace of the
extrinsic curvature of $S$ induced by its embedding in $\Sigma$, also
called the mean curvature. $H=0$ is then equivalent to $S$ having
extremal area, and $s^a H$ is the gradient of the area. In this case,
mean curvature flow is guaranteed to converge to a surface of $H=0$,
or extremal area, also called a minimal surface.  There is an extended
literature on mean curvature flow and minimal surfaces
\cite{Huisken}. Tod's idea is to generalize this method from
$H=\nabla_a s^a$ to $H=\nabla_a s^a-K+K_{ab}s^a s^b$.  For $K_{ab}\ne
0$, this flow is no longer guaranteed to converge, but one may hope
that it does, as the additional terms are of lower order.

One essential strength of generalized mean curvature flow is that it
cannot move a test surface {\it through} an AH, even for $K_{ab}\ne
0$. The argument is simple \cite{Bartnikpc}: Assume that the test
surface is about to move through the true AH, that is, it touches it
at one point. At that point both surfaces see the same $g_{ab}$,
$K_{ab}$ and $s^a$. Of the quantities which go into the expression
(\ref{H}) for $H$, only the $\nabla_a s^a$ differ on the two
surfaces. Keeping track of the signs, one sees that the test surface
must then always back away from the true AH at that point. Therefore,
a smooth test surface can never cross an AH (although it can approach
it asymptotically). This is true not only for generalized mean curvature
flow, but for all flows of the form $(\partial/\partial
\lambda)^a = - s^a \rho H$, as long as $\rho$ is strictly
positive. This property allows us to start the algorithm on a large
surface far out and evolve it inwards, thus making sure we find the
true AH.

We note that Eq. (\ref{Todeqn}) does not only specify the deformation
of $S$ as an abstract surface, but also identifies any point on $S$
with a point on its deformation. That information is not essential to
the method, and we get rid of it if we define $S$ as the level set
$F=0$. Then $s^a$ is again given by (\ref{s^a}). Consider a family of
moving surfaces $S(\lambda)$ given by $F(x^i,\lambda)=0$. On the
surface $F=const.$ we must have
\begin{equation}
{dF\over d\lambda} \equiv {\partial F\over \partial \lambda}
+ \left({\partial \over \partial \lambda}\right)^a \nabla_a F = 0,
\end{equation}
and therefore
\begin{equation}
\label{OSeqn}
{\partial F\over \partial \lambda} = - \left({\partial \over \partial
\lambda}\right)^a \nabla_a F = s^a H \nabla_a F = |\nabla  F|H.
\end{equation}
We note that (\ref{Todeqn}) is a geometric prescription: It specifies
a vector field on $S$ only in terms of the geometry of $S$ and
$\Sigma$ and the tensor field $K_{ab}$, independently of how $S$ is
parameterized. As we have just shown, the parabolic equation
(\ref{OSeqn}) is equivalent to it. We conclude that a flow
parameterized by, for example, $\partial F/\partial \lambda = H$,
without the factor $|\nabla F|$, does not have such a geometric
interpretation, but must depend on $F$ in a more general way than only
through the shape of its level set $S$. On the other hand, as $H$ is a
scalar function of $g_{ab}$, $s^a$ and $K_{ab}$ (evaluated on $S$), we
can replace it by any other scalar and still obtain a flow with
geometric meaning.  Any flow of the form
\begin{equation}
\label{flow}
{\partial F\over \partial \lambda} = |\nabla F| \times \text{any
 scalar}(K_{ab},g_{ab},s^a)
\end{equation}
is therefore geometric in nature. Such a general equation, replacing
$H$ by any function of the curvature of $S$, has already been given by
Osher and Sethian
\cite{OS}. 

If we now restrict $F$ to the form
$F(x^i,\lambda)=r-h(\theta,\varphi,\lambda)$, we have $\partial F/
\partial \lambda = - \partial h/ \partial \lambda$.  Therefore, any
flow of the form
\begin{equation}
\label{hflow}
{\partial h\over \partial \lambda} = -|\nabla (r-h)| \times \text{any
scalar}(K_{ab},g_{ab},s^a)
\end{equation}
is again geometric in nature. The naive Jacobi method,
Eq. (\ref{Jacobi}), however, is not.


\subsection{``Fast flow'' methods}


Before we propose our own AH finding algorithm, we summarize the
strengths and weaknesses of the existing ones. We have not discussed
algorithms which attack the elliptic problem directly via Newton's
method or a minimization iteration. Their main drawback, however, is a
small range of convergence, that is, they require a very good initial
guess. NKO is a lot more robust, but the need to treat $a_{00}$
separately is an important disadvantage. Eq. (\ref{a00}) is by no
means trivial: Solving it by any iterative method like Newton's method
is as computationally expensive as many steps of the main iteration
loop. Furthermore, for given data $(g_{ab},K_{ab})$ and given
$a_{lm}$, $l>0$, there may be several roots $a_{00}$ of this equation,
or none. Kemball and Bishop \cite{KB} propose investigating each root
of the equation separately, or if there are none, each minimum of the
right-hand side. Clearly this makes the algorithm even more expensive.

Most importantly, Newton's method for solving (\ref{a00}) tends to go
off into the wrong direction. As an example for this problem
\cite{Brandtpc}, we consider time-symmetric Cauchy data for the Kruskal
spacetime of mass $m$ in isotropic coordinates. [This is the special
case $m_2=0$ of Eq. (\ref{2bhdata}) below.] As a trial surface we take
a sphere of coordinate radius $\bar r$ centered on the black hole. The
expansion of outgoing null rays is
\begin{equation}
\label{H(r)eqn}
H = H(\bar r)={8\bar r(2\bar r-m)\over(2\bar r+m)^3}.
\end{equation}
From a mathematical point of view, this example is degenerate in the
sense that $H=0$ is a reduced from a differential to a purely
algebraic equation by the spherical symmetry of the trial
surface. (There is only one spatial direction, and this is the
degenerate direction of the elliptic operator.) Nevertheless, any
problems that arise in this toy equation also arise in a more
realistic situation. From a plot of $H(\bar r)$, Fig. 1, we see that
for $\bar r \gtrsim 1.87m$ Newton's method wanders off to infinity,
and for $r\lesssim 0.13m$ goes towards $\bar r=0$, instead of finding
the zero at $\bar r=m/2$. All algorithms which use Newton's method, or
a minimization method using derivatives, for any or all of the
$a_{lm}$, that is, the direct elliptic algorithms, share this problem.

The curvature flow method is sensitive only to the sign of $H$, not
its derivative. Applied to this problem, it goes towards smaller $r$
for positive $H$, and towards larger $r$ for negative $H$, and always
finds the apparent horizon. We have already seen that it cannot
accidentally walk through the AH. In these two properties lies its
robustness.  Any flow with $\rho H$ instead of $H$ on the right-hand
side, where $\rho$ is strictly positive and a scalar, shares the
fundamental advantages of the generalized mean curvature method: a
trial surface far outside the apparent horizon always moves in, and
can never accidentally cross the apparent horizon. As we have seen,
however, flow methods are slow because explicit discretizations in
time of parabolic methods require very small time steps for
stability. An implicit time step may be possible, but introduces a new
elliptic problem which a priori is not simpler than the underlying
elliptic problem one wants to solve.

We are not bound to geometrically motivated flows, however. Instead,
we heuristically consider all flow methods as variants of the Jacobi
method for solving $H=0$. Then we are free to combine the best
features of the curvature flow and NKO methods. From curvature flow we
would like to keep the properties that $a_{00}$ is not treated
specially, and that the change of $a_{00}$ should be proportional to
$H_{00}$, not $dH_{00}/da_{00}$.  From NKO we would like to adopt the
idea of subtracting and then inverting $L^2$, in order to suppress
high-frequency noise. This leads us to the following family of flows:
\begin{equation}
{\partial h\over \partial \lambda} = - A\left(1 - B L^2\right)^{-1}
\rho H,
\end{equation}
where $A$ and $B$ are free positive constants, and where $\rho$ is a
strictly positive weight depending on $h$ through at most first
derivatives.  The differential operator $1-BL^2$ is invertible, with
positive eigenvalues, for $B\ge 0$, and for $B>0$ its inverse is a
smoothing operator. When we discretize in $\lambda$, we can absorb
$\Delta\lambda$ into $A$. For simplicity we also restrict ourselves to
forward differencing. In spectral components, we obtain
\begin{equation}
\label{SpectralAHF}
a_{lm}^{(n+1)} - a_{lm}^{(n)} 
=
-{A\over 1 + B l(l+1)}
\left(\rho H\right)_{lm}^{(n)}.
\end{equation}
For $\rho$ we consider three choices: $\rho=1$ (``H flow''),
$\rho=|\nabla F|$ (``C flow''), and $\rho=|\nabla F| \sigma$ with
$\sigma$ defined in (\ref{rho}) (``N flow''). With $A>0$ and $B=0$, H
flow is the Jacobi method, and C flow is the curvature flow method. N
flow formally becomes the NKO method [compare
Eq. (\ref{NKOcomponents})] in the limit $A=B\to\infty$. The limit is
singular because the NKO method is not a flow and has to update the
component $a_{00}$ separately.

For determining the optimal values of $A$ and $B$, it is convenient to
reparameterize them with new parameters $\alpha$ and $\beta$ as
\begin{equation}
\label{alphabeta}
A = {\alpha\over l_{\text{max}}(l_{\text{max}}+1)} + \beta,
\quad
B = {\beta\over \alpha}.
\end{equation}
$A$ and $B$ now scale with $l_{\text{max}}$ in such a way that we
expect the optimal values of $\alpha$ and $\beta$ to be independent of
the value of $l_{\text{max}}$. $\alpha$ parameterizes an
$l$-independent contribution to the effective step size of
$\alpha/[l_{\text{max}}(l_{\text{max}}+1)]$, while $\beta$ adds an
$l$-dependent speedup which is zero at $l=l_{\text{max}}$ and
increases to $\beta$ at $l=0$.

It is clear that the fast flow methods have the potential to be much
faster than curvature flow, while still being numerically stable and
robust against bad initial guesses. They are not really flows of the
form (\ref{hflow}) because they are not local. In some situations, the
{\it effective} weight $\rho$ can become negative on parts of the
surface, and in these situations, the ``fast flow'' can move through
the true AH. Fast flow methods should be considered as (good)
compromises between the robustness of curvature flow and the speed of
NKO. Furthermore, one can trade robustness for speed by increasing
$\beta$, and vice versa, and so adapt the algorithm to the situation.
We have obtained the best results using N flow, with $\alpha=1.0$ and
$\beta=0.5$ in (\ref{alphabeta}). We note that $\alpha=1.0$, for any
$\beta$, means that the algorithm treats high frequency components
like the NKO algorithm does.


\section{Numerical implementation of pseudo-spectral apparent horizon finders}


The algorithm we suggest in this paper is formally defined by
Eq. (\ref{SpectralAHF}) with $\rho$ defined by (\ref{rho}) and
$A\simeq B\simeq 0.5$.

In order to implement this or any other pseudo-spectral algorithm, we
need to calculate the spectral components $(\rho H)_{lm}$ from the
spectral coefficients $a_{lm}$. In this section we give details of an
algorithm for doing this, given $g^{ij}$, $K_{ij}$ and $g_{ij,k}$ on a
Cartesian grid. We expect that there is scope for increasing the speed
and reducing the discretization error in this low-level part of the
algorithm, without changing the top-level part given by
(\ref{SpectralAHF}).


\subsection{The background structure}


The parameterization of the surface $S$ through spherical harmonics
and the introduction of the differential operator $L^2$ require the
introduction of a flat metric $\bar g_{ab}$. We do this by introducing
auxiliary Cartesian coordinates $\bar x^i=f^i(x^j)$, and then setting
the components of $\bar g_{ab}$ in the coordinates $\bar x^i$ to be
$\delta_{ij}$. The corresponding metric derivation $\bar\nabla_a$ is
then $\partial/\partial \bar x^i$, and $r^2=\delta_{ij}\bar x^i \bar x^j$. In
these coordinates $L^2$ is given by the expression
\begin{equation}
L^2 = (r^2 \delta^{ij}-\bar x^i\bar x^j){\partial\over\partial \bar x^i}
{\partial\over\partial \bar x^j} - 2\bar x^i{\partial\over\partial \bar x^i}.
\end{equation}
While more complicated choices are possible, we define
\begin{equation}
\bar x^i \equiv x^i - x^i_0,
\end{equation}
where $x^j$ are the Cartesian coordinates in which the Cauchy data are
presented to our algorithm. The freedom to shift the origin $r=0$
around is necessary because any trial surface will have to be
star-shaped around $r=0$, that is around $x^i=x^i_0$. Therefore we
have to make sure that $x^i=x^i_0$ is inside the AH.


\subsection{Calculating the $Y_{lm}$}
 

We need to calculate the $Y_{lm}(\bar x^i)$ and their first two partial
derivatives for arbitrary $(\bar x^i)$. Speed is important, because
our algorithm spends most of its time in these calculations.  The
standard spherical harmonics are
\begin{equation}
Y_l^m = \bar P_{l}^m(\cos\theta) e^{im\varphi}, 
\end{equation}
where the $\bar P_{l}^m$ are associated Legendre functions times a
constant depending on $l$ and $m$.  Instead of the complex $Y_{lm}$, we
introduce the real basis $\bar Y_{lm}$ as
\begin{equation}
\label{newbasis}
\bar Y_l^0 = \bar P_l^0(\cos\theta), \quad
\bar Y_l^{|m|} = \sqrt{2} \bar P_l^{|m|}(\cos\theta) \cos m \varphi, \quad
\bar Y_l^{-|m|} = \sqrt{2} \bar P_l^{|m|}(\cos\theta) \sin m \varphi.
\end{equation}
The real $\bar Y_{lm}$ obey the same conditions (\ref{cond0}-\ref{cond2})
as the standard complex $Y_{lm}$, but they are not eigenfunctions of
$L_z=\partial/\partial \varphi$. 
At each point $\bar x^i=(x,y,z)$ we calculate
\begin{equation}
\cos\theta=z/r, \quad \sin\theta=\rho/r, \quad \cos\varphi=x/\rho,
\quad \sin\varphi=y/\rho, 
\end{equation}
where
\begin{equation}
r=\sqrt{x^2+y^2+z^2},\quad\rho=\sqrt{x^2+y^2}.
\end{equation}
No explicit evaluation of trigonometric functions is required.
Then $\cos m\varphi$ and $\sin m\varphi$ are calculated as polynomials
in $\cos\varphi$ and $\sin\varphi$ from the recursion relations
\begin{eqnarray}
\cos m\varphi && = \cos (m-1)\varphi \, 
\cos\varphi - \sin (m-1)\varphi \, \sin\varphi, \cr
\sin m\varphi && = \cos (m-1)\varphi 
\, \sin\varphi + \sin (m-1)\varphi \, \cos\varphi. 
\end{eqnarray}
The $\bar P_{lm}$ are given explicitly for $m=l$ by 
\begin{equation}
\label{Zlm1}
\bar P_l^l(\cos\theta) = {1\over \sqrt{4\pi}}
{\sqrt{(2l+1)(2l)!} \over 2^l l!}
\,(-\sin\theta)^l,
\end{equation}
and for $0\le m\le l-1$ are calculated from the recursion relations
\begin{equation}
\label{Zlm2}
\bar P_{l}^m(\cos\theta) = \sqrt{2l+1\over l^2-m^2} \left[
\sqrt{2l-1} \,\cos\theta\,
\bar P_{l-1}^m(\cos\theta) - \sqrt{(l-1)^2-m^2\over 2l-3}
\bar P_{l-2}^m(\cos\theta)\right].
\end{equation}
(They are not needed for $m<0$.)  In order to calculate the first and
second partial derivatives with respect to $x$, $y$ and $z$, we
calculate the partial derivatives of $\bar Y_{lm}$ with respect to
$\theta$ and $\varphi$, and those of $\theta$ and $\varphi$ with
respect to $x$, $y$ and $z$, and explicitly code all terms arising
from the chain rule. The derivatives of $\bar P_{lm}(\cos\theta)$ with
respect to $\theta$ are obtained recursively after differentiating
Eqns. (\ref{Zlm1}) and (\ref{Zlm2}). The relations (\ref{cond1}) and
(\ref{cond2}) are then obeyed to machine precision by the numerically
calculated quantities.

We are aware of two other algorithms for calculating $Y_{lm}(\bar x^i)$
and their first and second partial derivatives. The algorithm of
Baumgarte {\it et al.} calculates them recursively as polynomials of
the $r_{,i}$. We have coded this algorithm directly from the detailed
formulae in \cite{Cornell}, and find that it scales in time as
$l_{\text{max}}^4$ and in storage requirement as
$l_{\text{max}}^3$. The NCSA/WashU apparent horizon finder
\cite{NCSA1,NCSA2} does not calculate the $Y_{lm}$, but a related basis
of smooth functions. This basis is not orthogonal, and it is not
independent of $r$ at constant $\theta$ and $\varphi$. For the
NCSA/WashU algorithm these properties of the basis functions are not
necessary. The calculation of this basis scales as approximately
$l_{\text{max}}^4$ in time, and as $l_{\text{max}}^4$ in storage
\cite{Anninospc}. 
In common with both algorithms, ours is recursive, and does not
require trigonometric function evaluations. The difference is that it
breaks up the $Y_{lm}$ into the product of a function of $\theta$ times
a function of $\varphi$. In consequence it scales $l_{\text{max}}^2$
in time (it is faster already for $l_{\text{max}}=2$), and as
$l_{\text{max}}^2$ in storage requirement.  This optimal scaling comes
at a price: the algorithm breaks down on the axis $x=y=0$, where
cancellations between the $\theta$ and $\varphi$ dependent factors in
the analytic expressions fail to take place numerically. In practice,
one can evade the problem by moving any collocation points that come
very close to the $z$ axis a small distance away from it, resulting in
a small error at that point, and a negligible one in the integrals
over $S$. Incorporating the cancellations into the code properly
requires mixing the $\theta$ and $\varphi$ dependency by going through
an intermediate, over-complete basis called ``symmetric trace-free
tensors'', which is precisely the approach of Baumgarte {\it et al.}


\subsection{Interpolating and integrating over $S$}


We need to discretize the integral $\int_S d\Omega$. We take as
collocation points on $S$ all those points where $S$ intersects a link
of the three-dimensional Cartesian grid. A link is the straight line
between two neighboring points on the Cartesian grid. The links which
intersect $S$ are those on which $F$ changes sign. The advantage of
this choice of collocation points is that one only needs to interpolate
in one dimension. Furthermore the number of collocation points on $S$
scales with the number of nearby points on the Cartesian grid, that
is, with the available numerical information.

To find the surface $F=0$ the algorithm calculates $F$ on all
Cartesian grid points. For this it needs the $\bar Y_{lm}$ on all grid
points.  Although these are required again and again, present
technology does not allow us to store
$l_{\text{max}}(l_{\text{max}}+1)$ 3D arrays for reasonable
$l_{\text{max}}$, so that they have to be recomputed each time.  Then
the algorithm flags all links on which $F$ changes sign. Both
operations scale as $N^3$, where $N$ is the linear grid size. We
determine by inverse linear interpolation where on the link the
intersection point is, then interpolate $g^{ij}$, $g_{ij,k}$ and
$K_{ij}$ to the intersection point by cubic interpolation. We
calculate $F_{,i}$ and $F_{,ij}$ directly at the $\bar x^i$ of the
intersection point. For this purpose we need the $r_{,i}$ and the
$\bar Y_{lm,i}$ and $\bar Y_{lm,j}$.

The integral $\int_S d \Omega$ is now approximated by the sum
\begin{equation}
\label{numintegral}
\int_S f d\Omega \simeq 4\pi {
\sum w  f
\over \sum w},
\end{equation}
where the sum is over all collocation points.  Let $\bar s^i$ be
the unit normal on $S$ with respect to the flat background metric $\bar
g_{ab}$,
\begin{equation}
\bar s^i = \left[\delta^{kl}(r_{,k}-h_{,k})(r_{,l}-h_{,l})\right]^{-1/2}
\delta^{ij}(r_{,j}-h_{,j}).
\end{equation}
The integration weight $w$ is then given by
\begin{equation}
w = {d\Omega\over dN} 
= {d\Omega\over dA} \,{dA\over dN}, \quad
{d\Omega\over dA} = {r_{,i} \bar s^i \over r^2}, \quad
{dN\over dA} =  
{|\bar s^1| \over \Delta \bar x^2\,\Delta \bar x^3} +{|\bar s^2| \over
\Delta \bar x^1\,\Delta \bar x^3} +{|\bar s^3| \over \Delta \bar
x^1\,\Delta \bar x^2},
\end{equation}
where $\Delta \bar x^i$ are the grid spacings on the three axes. Here
$d\Omega$ is the solid angle with respect to the flat metric $\bar
g_{ab}$ around $r=0$, that is $d\Omega = \sin\theta d\theta d\varphi$,
$dA$ is the surface element on $S$ induced by $\bar g_{ab}$, and $dN$
is the number of intersections of $dA$ with grid links.  Note that the
expression for $dN/dA$ models the anisotropy of the Cartesian grid in
an explicit sum over the three grid directions. The sum
(\ref{numintegral}) is a good approximation to the integral when
$\bar s^i$ changes little from one collocation point to its neighbors.

As a test of the discrete approximation to integrals over the surface,
we calculate the overlap integrals (\ref{cond0}) numerically. Let us
combine the indices $l$ and $m$ of the spherical harmonics into one
index $n$. The numerical approximation to the symmetric matrix $A_{nn'}
= \int Y_n Y_n'\,d\Omega$ is not exactly equal to the unit matrix,
because of the finite number of collocation points.

One could arrange the weights $w$ such that $A_{nn'}$ comes out right
for a given set of collocation points, but that would require putting
the collocation points in fixed, special positions with respect to
$\theta$ and $\varphi$, for example on a square grid in $\theta$ and
$\varphi$. In our algorithm, however, we let the position of the
collocation points be dictated by the underlying Cartesian grid $\bar
x^i$, and rely on a number of collocation points much larger than the
number $n_{\text{max}}=l_{\text{max}}(l_{\text{max}}+1)$ of basis
functions in order to keep the error down.  Table \ref{orthotable}
shows how the error in $A_{nn'}$ increases with $l$ for a surface $S$
created under realistic conditions by the apparent horizon finder. In
practice, the size of the 3-dimensional grids is limited by the
available computer storage, so that we have to choose $l_{\text{max}}$
small enough for the spectral error to remain small.

We can reduce this error in the following way. Let us denote
by $H_n=H_n(a_{n'})$ the true spectral components of the function $H$
on the surface parameterized by the expansion coefficients $a_n$. Let
$\tilde H_n$ be their numerical, slightly incorrect value. Clearly we
have
\begin{equation}
\tilde H_n = \sum_{n'=1}^\infty A_{nn'} H_{n'}.
\end{equation}
Without much additional numerical work, we can calculate a finite
square piece of the infinite matrix $A$ when we calculate $\tilde H_n$
up to $n_{\text{max}}$. Let $B$ be the inverse of that finite part of
$A$. Then we have (for $n\le n_{\text{max}}$)
\begin{equation}
\hat H_n \equiv \sum_{n'=1}^{n_{\text{max}}} B_{nn'} \tilde H_{n'}
= H_n + \sum_{n'=1}^{n_{\text{max}}} B_{nn'} 
\left(\sum_{n''=n_{\text{max}}+1}^\infty 
A_{n'n''} H_{n''}\right) \simeq H_n + \sum_{n'=n_{\text{max}}+1}^\infty 
A_{nn'} H_{n'}.
\end{equation}
In $\hat H_n$ the unwanted aliasing among the low ($n\le
n_{\text{max}}$) frequencies has been eliminated, and the remaining
deviation from the true value $H_n$ comes only from the aliasing of
high frequencies to low ones. One would assume this to be a better
approximation to $H_n$ than the $\tilde H_n$ in normal situations. In
practice, however, this assumption is difficult to test, as no cheap
estimate of the error $\hat H_n-H_n$ is available. 

Still, there are some indications of the remaining error in the $\hat
H_n$: The spectral components of $r$, the $r_n$, are by definition
identical to $a_n$. We find that the $\hat r_n$ are much closer to
$a_n$ than the $\tilde r_n$, but do not converge to them. The
remaining error can only be due to the fact that the collocation
points do not lie exactly on the true surface $S$ parameterized by the
$a_n$, due to the interpolation used to find them. This failure to
find the true surface $S$ is the only source of error for the $\hat
r_n$, but appears to be also the dominant source of error for the
$\hat H_n$, or any other nontrivial function on $S$.  In practice we
proceed as follows: We use $\hat H_n$ as our best approximation to
$H_n$. We monitor convergence of the final result $a_n$ of the AH
finder with $l_{\text{max}}$ and the grid spacing of the underlying
Cartesian grid.  We also monitor $|\hat r_n-a_n|$.  Finally, and
perhaps most importantly, we find that the algorithm using $\bar H_n$
converges better, and its error is considerably reduced when tested
against data for which the apparent horizon is known in closed
form. Therefore we always use the $\hat H_n$ and other hatted
quantities in the algorithm.

In order to estimate the quantity 
\begin{equation}
\int_S H^2 d\Omega =\sum_{n=1}^\infty H_n^2,
\end{equation}
which indicates to what precision the algorithm has found the apparent
horizon, we use the two numerically available quantities
\begin{equation}
\label{Hrms}
(H_{\text{rms}})^2 \equiv{\sum H^2 w \over \sum w} = \sum_{n=1}^\infty
\sum_{m=1}^\infty A_{nm} H_n H_m, \qquad\hbox{and}\qquad
|H|^2 \equiv \sum_{n=1}^{n_{\text{max}}} \hat H_n^2.
\end{equation}

After this work was carried out, we became aware of a different,
perhaps more efficient algorithm for going back between a function of
$\theta$ and $\phi$ and its spherical harmonic
components\cite{Norton}. One puts a grid on $S$ which is rectangular
and equally spaced in $\theta$ and $\varphi$, and then uses fast
Fourier transforms in $\theta$ and $\phi$. In a second step, one has
to discard those linear Fourier components which are not sufficiently
regular at the poles $\theta=0,\pi$, which is rather complicated. In
order to evaluate $g_{ij}$ etc. at the collocation points required
now, one has to interpolate in three dimensions from the given
Cartesian grid, instead of in one dimension. Nevertheless, there may
be scope for a more efficient algorithm here.


\section{Tests}


\subsection{Brill-Lindquist data}


The NCSA/WashU algorithm appears to be the only one to have been
tested on numerically evolved data \cite{NCSA2}. Tests that use data
given in closed form avoid interpolation from numerical data on a
grid, which poses an additional source of numerical error and even
instability in realistic situations. In the present paper we test our
algorithm with data given in closed form, but passed to the algorithm
only on the points of a numerical grid of realistic size. The input
into the code are the numerical values of the inverse metric and
extrinsic curvature components and of the first partial derivatives of
the metric components on the grid. A performance test with data
derived both from numerical initial data algorithms and numerical
time-evolutions is left to a future publication.

As a first test of the complete AH finder, we use Brill-Lindquist
time-symmetric initial data. For two black holes, these are
\begin{equation}
\label{2bhdata}
K_{ij}=0, \quad g_{ij} = \left(1+{m_1\over 2|x-x_1|}+{m_2 \over
2|x-x_2|}\right)^4 \delta_{ij}.
\end{equation}
The generalization to $N$ black holes is clear.  

We begin with a single black hole, where the AH is known explicitly:
it is a coordinate sphere of radius $m/2$.  There are a number of
possible convergence criteria for the iterative algorithm, none of
which fits all possible situations. One such criterium is
$H_{\text{rms}} > 2 |H|$. [These measures were defined in
(\ref{Hrms}).] This means that the residual of $H(\theta,\varphi)$ is
mainly in the high frequencies that we do not resolve. Table \ref{1bh}
shows the performance of the algorithm for this convergence
criterium. We chose a grid spacing such that there are roughly 16 grid
points across the interior of the AH. By the standards of a 3D single
grid numerical relativity code on a current supercomputer that is
already as much resolution as one can hope for. We chose
$l_{\text{max}}=6$, which is roughly the optimal value for that
resolution. The initial data are always $a_{00}=0.8$, while the horizon
radius is $0.5$. (For convenience, we use $a_{lm}$ rescaled by a factor
of $\sqrt{4\pi}$, so that $a_{00}$ is the average coordinate radius of
the AH.)

We have varied the offset of the center of the spherical harmonics
from the center of the AH. $\Delta r$ is the error in locating the AH
in coordinate space. It is calculated directly in those points where
the AH is collocated by the algorithm, not from the $a_{lm}$. The
result is roughly independent of the direction of the offset.  We see
that if the surface is very eccentric around the origin of
coordinates, precision suffers. Fortunately, there is a simple remedy:
If the dipole moments $a_1^{0,\pm1}$ are large, the algorithm
automatically uses them to obtain a better value for the origin
$x^i_0$ of coordinates, and restarts. This source of error is totally
eliminated by the procedure. The same remedy applies if the surface
touches the origin of coordinates at any stage during the flow.

The next test is Brill-Lindquist data for two uncharged black holes of
equal mass. We can position the two centers $x_1$ and $x_2$ so that
the metric is symmetric with respect to the $x=0$, $y=0$ and $z=0$
planes.  This allows us to work numerically on an octant of the full
grid, and save time and storage. The situation is in fact
axisymmetric, but the code does not know that. In
Fig. \ref{scatterplots} {\it all} points on the discretely represented
surface are plotted, giving coordinates $z$ versus
$\rho=\sqrt{x^2+y^2}$. The fact that they all fall on one curve shows
that the code represents an axisymmetric surface well in spite of the
underlying Cartesian grid.

In the data (\ref{2bhdata}) one can always find two minimal surfaces
surrounding $x_1$ and $x_2$. If $x_1$ and $x_2$ are close enough
together, there is a third minimal surface surrounding both of them.
Determining the maximal separation at which this happens is not an
easy test. Assume that the two centers are just far enough apart that
there no longer is a common horizon. By continuity there will still be
a smooth surface on which $H$ is small, but not zero,
everywhere. Numerically, this cannot be distinguished from a true
horizon.  

In the test, the two black holes have equal mass parameters
$m_1=m_2=1$. The total ADM mass is 2. We look for both inner and outer
surfaces. In Table
\ref{2bhtable} we show, for the same numerical parameters, the
root-mean-squared value of $H$ on the trial surface after our
algorithm has stopped, against the (coordinate) separation $d$ of the
two centers $x_1$ and $x_2$. The axisymmetric numerical algorithm of
Brill and Lindquist does not find an outer minimal surface for
$d>1.56$. From our calculations we can say with confidence that the
limit lies between $1.4$ and $1.6$. We should stress again that this
precision is limited by the resolution of the Cartesian grid on which
we give the Cauchy data. An algorithm specialized to axisymmetry could
of course determine this limit with much higher precision.


\subsection{Kerr data in Cartesian coordinates}


In order to test our code on analytic data with nonvanishing extrinsic
curvature, we consider Kerr data.  Cauchy data for the Kerr spacetime
have been given by Brandt and Seidel \cite{BS}. We transform these to
Cartesian coordinates by defining $x^ix^k\delta_{ij}=\bar r^2$, where
$\bar r$ is the radial coordinate which generalizes the isotropic
radial coordinate $\bar r$ for the Schwarzschild spacetime. For
testing our algorithm it is useful if we do not restrict the angular
momentum vector, or symmetry axis, to the $z$-axis, but give its
direction as a unit vector $n^i$, $n^in^j\delta_{ij}=1$. The
transformed expressions are
\begin{eqnarray}
g_{ij} && = A(\delta_{ij} + B v_i v_j), \quad
g^{ij} = A^{-1}\left(\delta^{ij} - {B\over 1 + v^k v_k B}
v^i v^j\right), \quad
K_{ij} = v_i w_j + w_i v_j, \\
v^i && = \epsilon^{ijk} n_j x_k, \quad
w^i = C x^i + D(n^i - \cos\theta x^i),
\end{eqnarray}
where $\cos\theta= n^k x_k / \bar r$, where all indices are moved with
$\delta_{ij}$, and where the coefficients are
\begin{eqnarray}
A && = \rho^2 \bar r^{-2}, \quad B = (\rho^2 + 2 m r) a^2 
\rho^{-2} \bar r^{-4}, \\
C && = \left[(r^2+a^2)^2-\Delta a^2\sin^2\theta\right]^{-1/2}
am \left[2r^2(r^2+a^2)+\rho^2(r^2-a^2)\right]\rho^{-3}\bar r^{-4}, \\
D && = \left[(r^2+a^2)^2-\Delta a^2\sin^2\theta\right]^{-1/2}
2a^3mr\Delta^{1/2}\cos\theta\rho^{-3}\bar r^{-4}, \\
\rho^2 && = r^2 + a^2 cos^2\theta, \quad \Delta = r^2 - 2mr + a^2,
\quad r=m + \bar r + {m^2-a^2\over 4\bar r}.
\end{eqnarray}
The apparent horizon is the coordinate sphere
$\bar r = \sqrt{m^2-a^2}/2$. For $a=0$ these data reduce to
Brill-Lindquist data for a single black hole. While $K_{ab}$ does not
vanish, the data are still special in that the two contributions
to $H$, $\nabla_a s^a$ and $m^{ab}K_{ab}$, vanish separately on the AH.
We have tested our algorithm on these data for different ratios of
$a/m$, different offsets $x^i_0$ between the center of the black hole
and the center of spherical harmonics, and for different orientations
$n^i$ of the black hole symmetry axis relative to the spherical
harmonics. The results are essentially the same as for the single
Brill-Lindquist black hole, giving an indication that the presence of
the extrinsic curvature term does not make a qualitative change to the
performance of the algorithm.


\section{Conclusions}


Numerical general relativity requires a fast and robust algorithm for
finding apparent horizons in Cauchy data without symmetries in three
dimensions given on a grid. In this paper we have described a new
apparent horizon finder algorithm which appears to be as fast but more
robust than its best predecessor.

We began from a general classification of possible approaches to the
problem. Any approach which poses the problem as a nonlinear elliptic
equation on a topological two-sphere and then attacks that equation
directly will fail unless provided with a very good initial guess,
because the problem is nonlocal in nature. While we have disregarded
such approaches here, they will be ideal as a second stage whenever
the apparent horizon needs to be determined to high precision.  We
concluded that for robustness the best algorithm is probably the
(generalized) mean curvature flow suggested by Tod \cite{Tod}, where
an arbitrary initial surface evolves in an unphysical ``time'' towards
the apparent horizon, turning the problem from an elliptic into a
parabolic one.  The algorithm is guaranteed to converge at least for
time-symmetric ($K_{ab}=0$) data, and we have argued that it must be
at least very robust also for $K_{ab}\ne0$ data. Unfortunately,
numerical implementations of this algorithm face a numerical stability
problem common to all parabolic equations, which make them slow, and
increasingly so with increasing resolution, in practice.

This stability or speed problem is not present in the algorithm of
Nakamura {\it et al.} (NKO) \cite{NKO}. It is motivated by a standard
way of solving nonlinear elliptic problems numerically, namely
subtracting a simple linear elliptic operator from the nonlinear one,
inverting it by pseudo-spectral methods and iterating.  Here we have
thrown more light on how NKO works, by making explicit the background
metric it introduces, and by characterizing the iteration procedure as
a specific finite differencing, in unphysical time, of a parabolic
problem.  This parabolic problem itself is the singular limit of a
certain family of flows which are governed by a mixture of the
physical geometry of the Cauchy data and an unphysical background
geometry. Tod's flow is a different limiting case of that family, one
in which no background metric appears.

Once we have recognized the existence of a continuum of possible
algorithms between Tod and NKO, it is plausible that an algorithm
somewhere in the middle of the continuum may be better than the
extremes. By trial and error, we have determined the optimal member of
the family of algorithms. This intermediate algorithm evolves the
high-frequency components (the fine details) of the trial AH
essentially like the NKO algorithm, but it evolves the low frequency
components (the rough shape) by a variant of generalized mean
curvature flow. We therefore call it "fast flow".

We have given details of a numerical implementation of the
pseudo-spectral methods which are needed for implementing both the
original NKO and our new algorithm. Such details have not been
published before. It should be stressed that the formal analysis of
the algorithm in Sect. III is independent from its implementation in
Sect. IV, and there may be different and more efficient
implementations.

We have not made direct performance comparisons with other
algorithms, and the tests we have described are viability rather than
performance tests. Nevertheless, we anticipate the following:

By construction, the new algorithm is as fast as that of NKO: The
iteration steps are very similar, and there is the same small number
of them. NKO, however, updates $a_{00}$ (the overall radius of the
trial AH) by a special procedure.  According to how this is done
\cite{KB}, the additional overhead may be large. More
importantly, the separate update of $a_{00}$ has the potential to
reduce the robustness of NKO: Eqn. (\ref{a00}) may have several
solutions, in which case all should be investigated, or none, in which
case minima should be investigated instead \cite{KB}. This requires
some decision-taking, which will be hard to automate, or instead an
infinitely branching search. We have also argued that zeros of eqn.
(\ref{a00}) are hard to find.  Either NKO or fast flow should be far
more tolerant of initial guesses than the elliptic methods.

The method of choice for robustness and elegance is clearly Tod's mean
curvature flow. The only question here is speed. We have argued that
as a parabolic method this would be slow in the possible
implementations known to us, but a quantitative comparison with the
implementation of Pasch \cite{Pasch} would be interesting.

The only existing algorithm directly accessible to the author is that of the
NCSA/WashU group \cite{NCSA1,NCSA2}. Direct comparisons are planned in
future realistic applications. The new algorithm seems to be more
robust, though: it made the transition from two AHs to a single one in
the family of Brill-Lindquist data summarized in Table 1 without
external input. This ability will be crucial for applying AH boundary
conditions in the merger of two black holes when the computer code is
on its own during a very large run.  The NCSA/WashU code has not been
tested on a similar sequence of analytic initial data, but in some
situations involving evolved black holes it presently requires some
care in finding the correct horizon\cite{CSpc}.

Finally, the source code of the new spectral AH finder will be
published early in 1998 in conjunction with the "Cactus" numerical
relativity infrastructure \cite{Cactus}.

        
\acknowledgments

I am grateful to Bernd Bruegmann and Steve Brandt for helpful
discussions on all aspects of this paper, and to Ed Seidel for a
critical reading.



\begin{table}
\caption{Maximal deviation of the overlap matrix $A_{nn'}$ from the
unit matrix as a function of the linear grid size, the integration
surface $S$ and $l_{\text{max}}$}.
\label{orthotable}
\begin{tabular}{llddd}
Surface parameters & grid size &
$l_{\text{max}}=4$ & $l_{\text{max}}=8$ & $l_{\text{max}}=16$ \\ 
\tableline
$a_{00}=1.0$ & 16 & 0.053 & 0.159 & 0.284 \\
& 32 & 0.016 & 0.037 & 0.069 \\
& 64 & 0.011 & 0.026 & 0.040 \\ 
$a_{00}=1.0$, $a_{1,-1}=0.4$ & 32 & 0.039 & 0.069 & 0.178 \\
& 64 & 0.047  & 0.047 & 0.053 \\
$a_{00}=1.0$, $a_{10}=0.4$ & 32 & 0.052 & 0.087 & 0.274 \\
& 64 & 0.050 & 0.056 & 0.075 \\
\end{tabular}
\end{table}


\begin{table}
\caption{Root-mean-square residuals of $H$ and maximal errors in the
position of the numerically calculated AH, for Schwarzschild data
offset from the coordinate origin. The coordinate radius of the AH is
$0.5$.}
\label{1bh}
\begin{tabular}{drrr}
Offset & iterations & $H_{rms}$ & $(\Delta r)_{max}$ \\
\tableline
0.0 & 10 & $9\times 10^{-4}$  & $7\times 10^{-4}$ \\
0.1 & 10 & $9\times 10^{-4}$  & $7\times 10^{-4}$ \\
0.2 & 11 & $1\times 10^{-3}$  & $8\times 10^{-4}$ \\
0.3 & 12 & $2\times 10^{-3}$  & $1\times 10^{-3}$ \\
0.4 & 10 & $6\times 10^{-2}$  & $2\times 10^{-2}$ \\
\end{tabular}
\end{table}


\begin{table}
\caption{Root-mean-square residuals of $H$ on the inner and outer
numerically calculated minimal surface in Brill-Lindquist data for two
black holes of equal mass. ``No convergence'' is the unaided return
status of the algorithm. It means that the residual value of $H$ given
in brackets is not due to a lack of numerical resolution.}
\label{2bhtable}
\begin{tabular}{drr}
Separation & inner $H_{\text{rms}}$ & outer $H_{\text{rms}}$ \\
\tableline
0.0 & only one surface & $1.9\times 10^{-5}$ \\
0.4 & $8 \times 10^{-2}$  & $1.8 \times 10^{-5}$ \\
0.8 & $9 \times 10^{-5}$ & $1.8 \times 10^{-5}$ \\
1.2 & $3.0 \times 10^{-4}$ & $1.5 \times 10^{-4}$ \\
1.4 & $2.6 \times 10^{-4}$ & $2.0 \times 10^{-3}$ \\
1.6 & $2.8 \times 10^{-4}$ & no convergence ($3.0 \times 10^{-2}$) \\
1.8 & $2.4 \times 10^{-4}$ & no convergence ($3.9 \times 10^{-1}$) \\
2.0 & $7 \times 10^{-4}$ & (not attempted) \\ 
\end{tabular}
\end{table}


\begin{figure}
\centerline{\epsffile{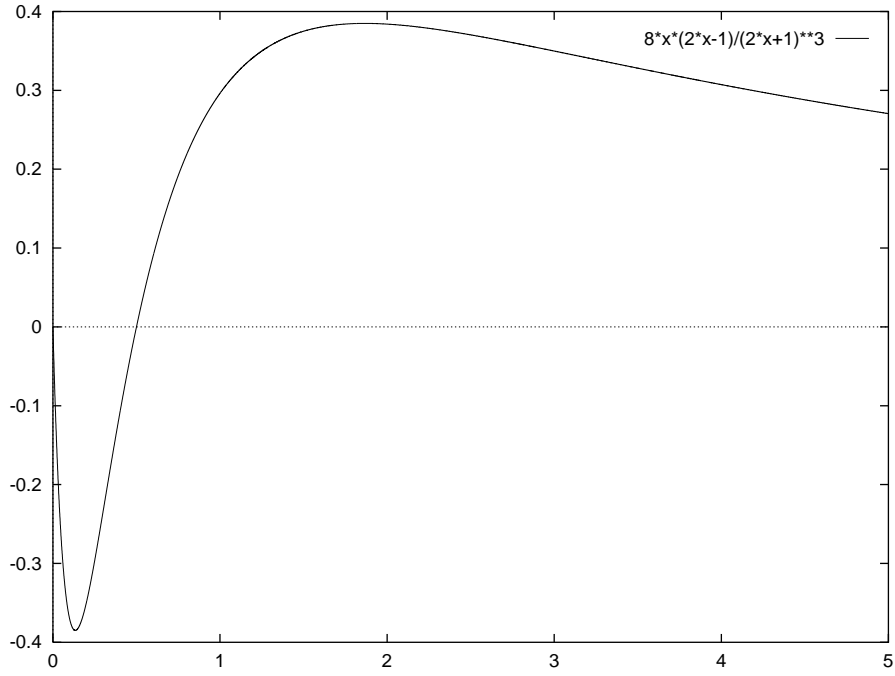}}
\label{H(r)plot}
\caption{Plot of the horizon function $H(\bar r)$ versus $\bar r$, in units of
the black hole mass $m$, as given in Eq. (\ref{H(r)eqn}).}
\end{figure}


\begin{figure}
\centerline{\epsffile{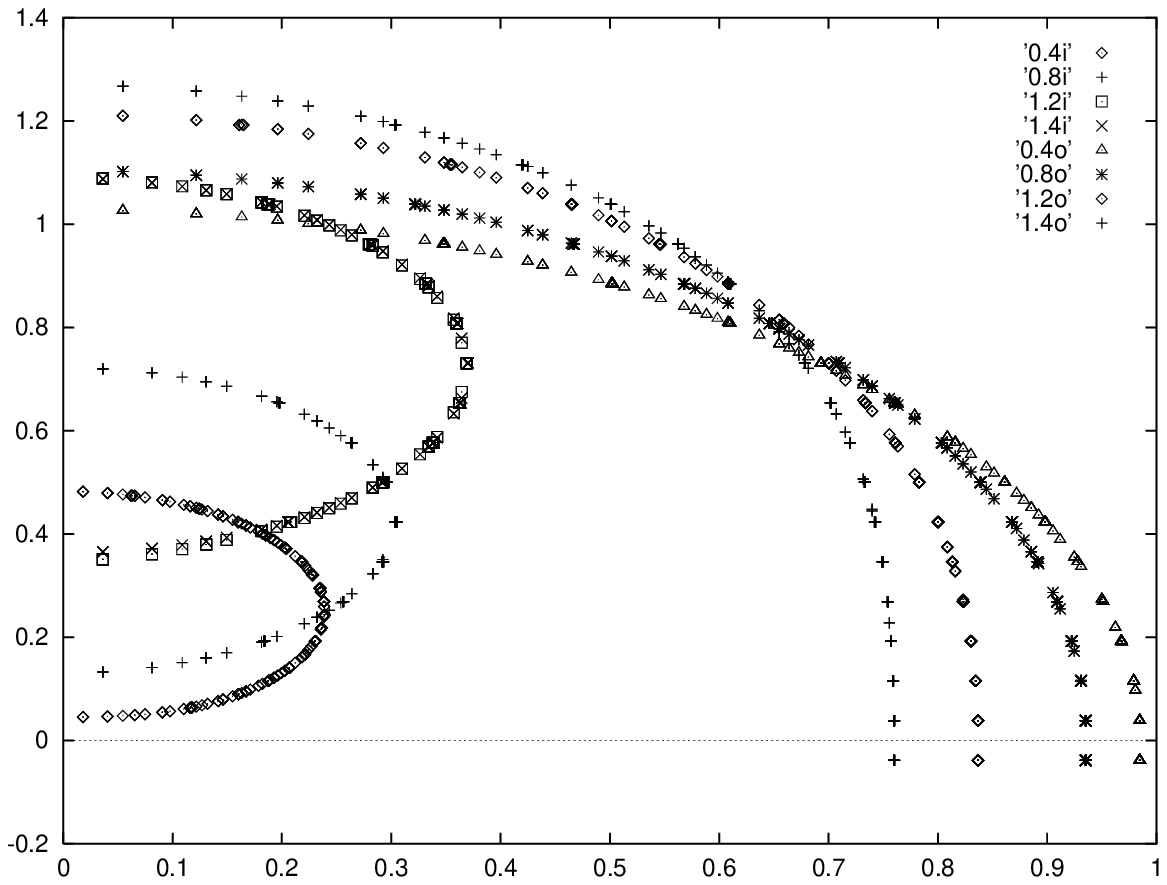}}
\label{scatterplots}
\caption{Shape of the AH in the axisymmetric, z-reflection-symmetric
situation. The algorithm assumes x, y and z-reflection symmetry, but
not axisymmetry. The plot shows $z$ versus $\sqrt{x^2+y^2}$ for all
grid points on the AH in one octant of the full grid. The small half
circles are the inner horizons, for a separation of $d=0.4$, $0.8$,
$1.2$ and $1.4$ of the two black holes, from bottom to top. The large
quarter circles are the outer horizons, from right to left, or bottom
to top.}
\end{figure}



\begin{references}
\bibitem{Anninospc} I thank Peter Anninos for making his Fortran code
for the calculation of ``symmetric trace-free tensors'' available.
The scaling with $l_{\text{max}}$ can be inferred already from the
papers \cite{NCSA1,NCSA2}.
\end{references}
\end{document}